\documentclass[aps,pra,10pt,superscriptaddress,twocolumn]{revtex4}
\usepackage{times,amsmath,amssymb,amsfonts,graphicx,bbm}
\usepackage{lineno,color}

\begin{document}
\title{Quantum metrology out of equilibrium}
\author{Sholeh Razavian}
\affiliation{Faculty of Physics, Azarbaijan Shahid Madani University, Tabriz, Iran}
\author{Matteo G. A. Paris}
\affiliation{Quantum Technology Lab, Dipartimento di Fisica {\it 'Aldo Pontremoli'}, 
Universit\`a degli Studi di Milano, I-20133 Milano, Italy}
\date{\today}
\begin{abstract}
We address open quantum systems out-of-equilibrium 
as effective quantum probes for the characterisation of their 
environment. We discuss estimation schemes for parameters 
driving a de-phasing evolution of the probe and then focus 
on qubits, establishing a relationship between the quantum Fisher 
information and the residual coherence of the probe. Finally, we apply
our results to the characterisation of the ohmicity parameter of 
a bosonic environment.
\end{abstract}
\maketitle
\section{Introduction}
\begin{figure}[h!]
\includegraphics[width=0.98\columnwidth]{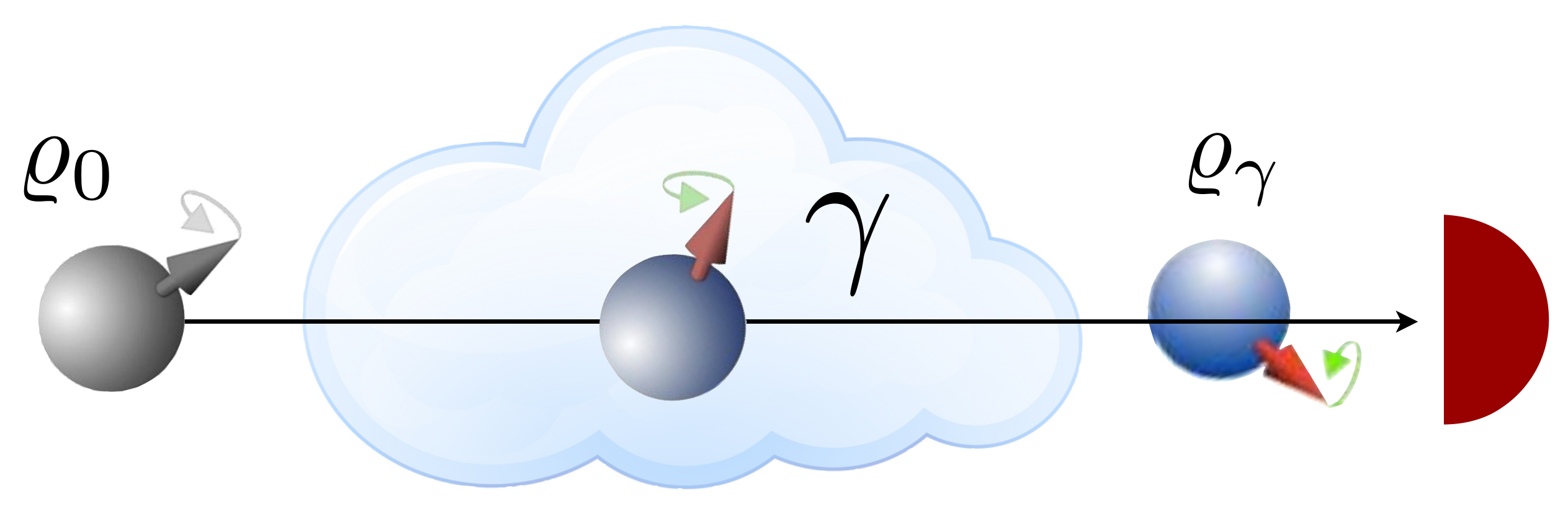}
\caption{A quantum probe is prepared in a known initial 
state and then made to interact with a larger, and possibly complex, 
system, which represents the environment of the probe. The 
output state of the probe thus carries information about the 
properties of the environment, e.g. the parameter $\gamma$, which
may be extracted by performing measurements at the output.}\label{f:sch}
\end{figure}

In this paper we address open quantum systems 
out-of-equilibrium \cite{qout1,qout2}, employed 
as quantum probes to precisely characterise some relevant 
properties of their environment \cite{elliott,streif16,cosco17}. The probing scheme 
we are going to discuss
is the following (see Fig. \ref{f:sch}): a quantum probe, i.e. a 
simple quantum system like a qubit, is prepared in a known initial 
state and then made to interact with a larger, and possibly complex, 
system, which represents the environment of the quantum probe. The 
environment usually induces decoherence, to an amount which depends 
on its temperature, spectral density and internal correlations \cite{palma96}. The 
final state of the quantum probes thus carries information about the 
properties of the environment. In turn, any measurement performed 
on the probe may be exploited to infer the values of some relevant 
environment parameters \cite{benedetti14,p14,zwick16,fujiwara03,pinel13}. In this situation, the inherent fragility 
of quantum systems to decoherence represents a resource, making 
quantum probes a very effective technique, able to provide enhanced 
precision compared to classical (thermal) probes \cite{brunelli,brunelli2}. 
In addition, quantum
probes are usually {\it small} and do not perturb the system under 
investigation, thus representing a non-invasive technique suitable 
for delicate samples \cite{Qthrmo,horodecki}. 
\par
The interaction time between the probe and the system is usually a 
tunable parameter, and a question thus arises on whether it may be 
used to further optimise the estimation precision. In a classical 
setting, e.g. thermometry, this is not the case, since one prepares 
the probe, leave it interacting with the sample, and then read the 
environment parameter by measuring the probe when it has reached its 
stationary state, i.e. it is at equilibrium with its environment.
On the other hand, it has been recently shown that optimal estimation by quantum probes may be achieved also at finite 
time, i.e. when the probe has not reached stationarity, and 
it is still in an out-of-equlibrium state \cite{J1D,m14,r18,c18,adesso,jevtic,geno18}. 
Following this results, we address here 
open quantum systems out-of-equilibrium as possible quantum probes 
for the characterisation of their environment. At first, we discuss 
a general scheme to estimate parameters driving the de-phasing 
evolution of a probe. We then devote attention to qubit probes, and 
establish a relationship between the quantum Fisher information 
and the residual coherence of the probe. 
\par
The paper is structured as follows. In Section 2, we establish 
notation, describe the dynamics of a probe subject to dephasing,
and introduce the notion of residual coherence. In Section 3, we 
briefly review the tools of quantum parameter estimation. 
In Section 4, we address quantum probes out-of-equilibrium,  
establish a relationship between the quantum Fisher 
information and the residual coherence of the probe, and  
apply our results to the characterisation of the ohmicity 
parameter of a bosonic environment. Section 5 closes the paper
with some concluding remarks.
\section{Probing by dephasing}\label{s:pbd}
Let us consider a generic quantum system interacting with its
environment. No assumptions is made on the dimension $d$ of the 
Hilbert space of the probe. We also do not assume any specific
form for the interaction Hamiltonian, but nevertheless assume 
that the resulting dynamics corresponds to a pure dephasing \cite{s18,cp14,mact,addis}, i.e. to a Von Neumann-Liouville 
equation of the form
\begin{align}
\dot \varrho & = - i\, [H,\varrho] - \kappa \, [H,[H,\varrho]\,] \\
& = - i\, [H,\varrho] + 2\,\kappa\, L[H]\, \varrho\,,
\end{align}
where $\varrho$ is the density matrix describing the state of the 
system, $H$ is its free Hamiltonian, $\kappa >0$ is a dephasing rate
and $L[O]\,\bullet = - \frac12 \{O^\dag O, \bullet\} + O \bullet O^\dag$
is a super-operator in the Lindblad form.
Moving to the interaction picture, i.e. to a reference frame rotating 
with $H$, the equation of motion reduces to
\begin{align}
\dot \varrho = 2\,\kappa\, L[H]\, \varrho\,.
\end{align}
Upon writing the initial state in the Hamiltonian basis, i.e.  
$\varrho_0 = \sum_{nk} \varrho_{nk} |e_n\rangle\langle e_k|$ with
$H |e_n\rangle = E_n |e_n\rangle$, the state after the interaction with
the environment is given by 
\begin{align}\label{rhf}
\varrho_\gamma & = \sum_{nk} \varrho_{nk}\, e^{-\gamma\, \Omega_{nk}^2}\,
|e_n\rangle\langle e_k|\,, \\
& = \int_{\mathbb R}\! dz\, g(z;0,2\gamma)\, e^{-i z H} \varrho_0\, e^{i z H}
\end{align}
where $\gamma = \kappa t$, $\Omega_{nk} = E_n - E_k$ and $g(z;\bar z,\sigma^2)$ is a Gaussian distribution in the variable 
$z$ with mean $\bar z$ and variance $\sigma^2$.  
The coherence of the 
probe after the interaction is given by
\begin{align}
C_\gamma = \sum_{n\neq k} |\varrho_{nk}|\, e^{-\gamma\, \Omega_{nk}^2} = 
2 \sum_{n < k} |\varrho_{nk}|\, e^{-\gamma\, \Omega_{nk}^2}\,,
\end{align}
and is always smaller than the initial coherence $C_0 = 2 \sum_{n < k} 
|\varrho_{nk}|$.
Since the precision of quantum probes is strictly related to their
sensitivity to decoherence, it is quite natural to start from a probe
initially prepared in a maximally coherent state, i.e. $\varrho_0 =
|\psi_0\rangle\langle\psi_0|$ where
\begin{align}
|\psi_0\rangle = \frac{1}{\sqrt{d}} \sum_{nk} |e_n\rangle\langle e_k| 
\qquad C_0 = 1\,.
\end{align}
For the sake of simplicity we also assume equi-spaced levels for the probe,
i.e. $\Omega_{nk}^2 = \Omega^2 (n-k)^2$. In this case, the residual coherence 
after the interaction may be expressed as
\begin{align}
C_\gamma = \frac2d \sum_{j=1}^{d-1} e^{-j^2\,\gamma\, \Omega^2}\,.
\end{align}
\section{Quantum parameter estimation}
After the interaction with the environment the state of the
probe depends on the parameter we would like to estimate. In 
order to optimize the inference strategy, i.e. to optimize 
the extraction of information, we employ the tools of quantum 
estimation theory \cite{Helstrom,parisQT}, which provides recipes to 
find the best detection scheme and to evaluate the corresponding 
lower bounds to precision. The precision also depends on the 
interaction time or, equivalently, on the residual coherence 
of the probe.
\par
Let us consider the family of quantum states $\rho_\gamma$ and
assume that the dephasing rate depends on some parameter
of interest, i.e. $\gamma = \gamma (\lambda)$. 
We perform measurements on repeated preparations of the probe 
and then process the overall sample of outcomes in order to
estimate $\lambda$. Let us denote by $Z$ the observable we 
measure on the probe ($Z|z\rangle = z |z\rangle$, $P_z=|z\rangle\langle z|$), 
and by $p(z|\lambda)=\hbox{Tr}[\varrho_{\gamma(\lambda)}\,P_z]$ 
the distribution of its 
outcomes for a given value of $\lambda$. After choosing a certain
observable $Z$, we perform $M$ repeated measurements, collecting the 
data ${\mathbf z}=\{z_1,...,z_M\}$. This set is then processed  
by an {\it estimator} $\widehat{\lambda} \equiv \widehat 
\lambda ({\mathbf z})$, i.e. a function from the space of data to 
the set of possible values of the parameter. The {\it estimate} 
value of the parameter is the mean
value of the estimator over data, i.e.
\begin{align}
\overline{\lambda} = \int\!\! d{\mathbf z}\, p(\mathbf{z}|\lambda) 
\, \widehat\lambda ({\mathbf z})\,,
\end{align}
where $ p(\mathbf{z}|\lambda) = \Pi_{k=1}^M\, p(z_k|\lambda)$ since 
the repeated measurements are independent on each other. The 
{\it precision} of this estimation strategy corresponds
to the variance of the estimator i.e. 
\begin{align}
V_\lambda \equiv \hbox{Var}\, \lambda = 
\int\!\! d{\mathbf z}\, p(\mathbf{z}|\lambda) 
\, \Big[\widehat\lambda ({\mathbf z})- \overline{\lambda} \Big]^2\,.
\end{align}
The smaller is $V_\lambda$, the more precise is the estimation 
strategy. In fact, the precision of any unbiased estimator (i.e. an estimator
such that $\overline{\lambda} \rightarrow \lambda$ for $M \gg1$), 
is bounded by the so-called Cram\`er-Rao (CR) inequality:
\begin{equation}
V_\lambda \ge \frac{1}{M F_\lambda}
\end{equation}
where $F_\lambda$ is the Fisher information (FI) of $Z$
\begin{align}
F_\lambda = \int\!\! dz\, p(z|\lambda) \,\Big[\partial_\lambda 
\log p(z|\lambda)\Big]^2\,,
\end{align}
i.e. the information that can be extracted on $\lambda$ by performing 
measurements of $Z$ on $\varrho_{\gamma(\lambda)}$.
The best, i.e. more precise, measurement to infer the value of 
$\lambda$ is the measurement maximising the FI, where the maximization
is performed over all the possible probe observables. 
\par
As a matter of fact, the maximum is achieved for any observable 
having the same spectral measure of the so-called {\it symmetric logarithmic derivative} $L_\lambda$, i.e. the selfadjoint operator satisfying 
the equation
\begin{equation}
2\,\partial_\lambda\varrho_{\gamma(\lambda)} = L_\lambda\,\varrho_{\gamma(\lambda)} +\varrho_{\gamma(\lambda)}\, L_\lambda\,.
\end{equation}
The corresponding FI is usually referred to as the {\it quantum Fisher information}
(QFI) and may be expressed as $H_\lambda = \hbox{Tr}[\varrho_{\gamma(\lambda)}\,L_\lambda^2]$. Since $F_\lambda \leq H_\lambda$, the ultimate bound to precision
in estimating $\lambda$ by performing quantum measurements on $\varrho_{\gamma(\lambda)}$ is given by the quantum CR bound
\begin{equation}
V_\lambda \ge \frac{1}{M H_\lambda}\,.
\end{equation}
In terms of the eigenvalues and eigenvectors of 
$\varrho_\gamma= \sum_n \varrho_{n} |\phi_{n}\rangle\langle\phi_{n}|$ the 
QFI may be written as
\begin{equation}\label{qfif}
H_\lambda= \sum_n \frac{(\partial_\lambda\varrho_n)^2}{\varrho_n}+2 \sum_{n \ne k}\frac{(\varrho_n-\varrho_k)^2}{\varrho_n+\varrho_k}\,\big|\langle\phi_k|\partial_\lambda \phi_n\rangle\big|^2\,.
\end{equation}
In order to evaluate analytically the QFI, we need to diagonalise the
density matrix of the probe after the interaction, i.e. that in Eq. 
(\ref{rhf}). This can be done easily  for low dimensional probes 
(qubit and qutrits), whereas numerical solutions 
are often needed for higher dimensions \cite{hf14}. 
Upon maximising the QFI one then
optimises the estimation scheme \cite{ps12, ps16}. 
\par
A global measure of the estimability of a parameter, which 
compares the variance with the value of the parameter, is given 
by the signal-to-noise ratio $R_\lambda=\lambda^2/V_\lambda$.
In turn, the quantum CR bound may be rewritten in terms of 
$R_\lambda$ as follows
\begin{align}\label{qsnr}
R_\lambda \leq Q_\lambda = \lambda^2 H_\lambda\,,
\end{align}
where $Q_\lambda$ is referred to as the quantum signal-to-noise 
ratio (QSNR).
\section{Quantum probes out of equilibrium}
Let us now consider the simplest quantum probe, i.e. a qubit system used 
to characterize its environment, which itself induces dephasing on the 
qubit. In this case we may consider a generic pure initial state
$|\psi_0\rangle = \cos\phi |e_1\rangle + \sin\phi |e_2\rangle$
and use the notation $\Omega=\Omega_{21} = E_2- E_1$. The initial coherence 
is given by $C_0=\sin 2\phi$ and the final one by $C_\lambda 
\equiv C_{\gamma(\lambda)}= e^{-\gamma(\lambda) \Omega^2} \sin 2\phi$. In order to 
evaluate the QFI, we leave the qubit to evolve, then 
diagonalise the state, and finally use Eq. (\ref{qfif}). After some 
algebra, we obtain a remarkably compact formula 
\begin{align}\label{cqfi}
H_\lambda = \Omega^4\, \big(\partial_\lambda \gamma\big)^2 \frac{C_0^2 C_\lambda^2}{C_0^2 - C_\lambda^2}\,,
\end{align}
which is valid for any $\phi$ and expresses the QFI in terms of the dependence 
of the dephasing rate on the parameter of interest, i.e. the {\it susceptibility} 
$\partial_\lambda \gamma$, and on the relationship between the initial and the final
coherence of the probe. As it may easily proved, 
the maximum of the QFI is achieved for $\phi=\pi/4$, thus
confirming the intuition, already mentioned in Section \ref{s:pbd}, that the 
optimal initial state of the probe corresponds to a maximally coherent state 
$|+\rangle = (|e_1\rangle + |e_2\rangle)/\sqrt{2}$.
We remark that Eq. (\ref{cqfi}) is valid for any (pure) initial 
preparation of the probe and any kind of parameter, the only 
assumption being that the preparation of the environment and 
the corresponding interaction leads to a pure dephasing 
evolution of the probe.
\begin{figure}[h!]
\includegraphics[width=0.98\columnwidth]{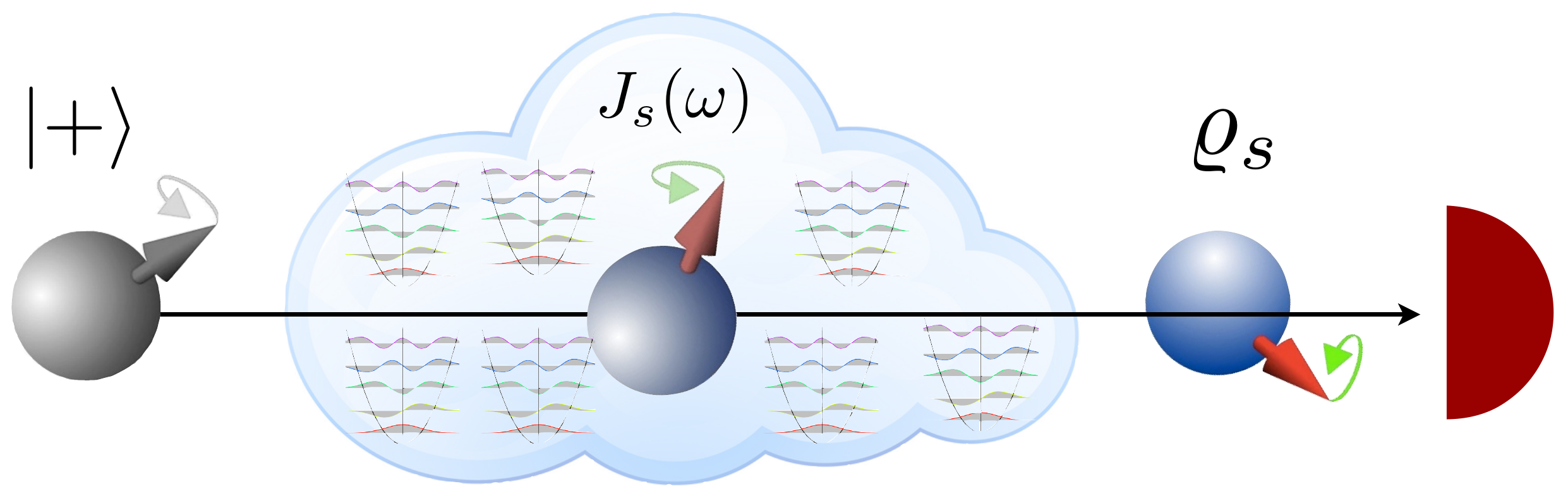}
\caption{A qubit is prepared in a maximally coherent state 
$|+\rangle = (|e_1\rangle + |e_2\rangle)/\sqrt{2}$ and then is 
made to interact with a Ohmic-like environment made of bosonic 
modes, characterised by a spectral density $J_s(\omega)$
The output state of the probe carries information about the 
ohmicity parameter $s$, which may be estimated by performing 
measurements at the output.}\label{f:sc2}
\end{figure}
\par
As an application, let us now consider a qubit, used to probe the nature
of an Ohmic-like environment made of bosonic modes (see Fig. \ref{f:sc2}). 
In order to introduce the problem, let us write the Hamiltonian of 
the whole system. We use the natural system of units ($\hbar=1$), 
and also write the Hamiltonian in unit of the qubit frequency 
$\Omega$, making it adimensional
\begin{equation}\label{htot}
{\cal H}=\frac12 \sigma_{3} +\sum_{k}\omega_{k}\,b^{\dagger}_{k}
\,b_{k}+\sigma_{3}\, \sum_{k}(g_{k}\,b^{\dagger}_{k}+g^{*}_{k}\,b_{k})\,,
\end{equation}
where $\omega_{k}$ is the (dimensionless) frequency of the $k$-th 
environmental mode. The $\sigma$'s are the Pauli matrices and
$[b_{k},b^{\dagger}_{k}]=\delta_{k\,k^{'}}$ describe the modes
of the environment. The $g_{k}$'s are coupling constants, describing 
the interaction of each mode with the qubit probe. Their distribution 
determines the {\it spectral density} of the environment, according to
the expression $J(\omega)=\sum_{k}\,|g_{k}|^{2}\,\delta(\omega_{k}-\omega)$. 
The spectral density is the crucial quantity to describe the system-environment 
interaction, and it does depend on the specific features of the environment. 
In turn, the characterisation of the spectral density is 
crucial to understand, and possibly control, quantum 
decoherence\cite{Paavola,Martinazzo,Myatt,
PiiloManiscalco,Grasselli,claudia}. 
\par 
A large class of structured reservoirs is characterised by an 
{\it Ohmic}-like spectral density of the form
\begin{equation}
J_s(\omega)= \omega_c \left(\frac{\omega}{\omega_c}\right)^s 
\exp\left\{-\frac{\omega}{\omega_c}\right\}\,,
\end{equation}
where the frequencies are in unit of $\Omega$.
The cutoff frequency describes a natural boundary in 
frequency response of the system. As we will see, it determines the
timescale of the evolution. The quantity $s$ is a real 
positive number, which governs the behaviour of
the spectral density at low frequencies. Upon varying $s$ we move 
from the so-called sub-Ohmic regime ($s<1$), to Ohmic ($s=1$), and 
to super-Ohmic one ($s>1$). Different values of the ohmicity parameter 
$s$ often corresponds to radically different kinds of dynamics, and 
therefore it would be highly desirable to have an estimation scheme 
for their precise characterisation.
\par
Such a scheme may be obtained from the results of the previous Sections,
since the dynamics induced on the qubit is a pure dephasing. In order 
to prove this result, one assumes that the system is initially
in the state $|\psi\rangle \otimes |0\rangle$ (i.e. a generic state for the probe and, assuming to be at zero temperature, 
the ground state for the enviroment), then evolve the 
whole system according to the Hamiltonian in Eq. (\ref{htot}), 
and finally trace out the environment. The resulting
evolution is that of Eq. (\ref{rhf}) where the dephasing rate 
is given by 
\begin{align}
\gamma_s (\tau) & =\int_0^{\infty}  \frac{1-\cos(\omega \tau/\omega_c)}
{\omega^2}\,J_s(\omega)\, d\omega\,, 
\\
& \notag
\\
& = \left\{\begin{array}{ll}
\frac12 \log\left(1+\tau^2\right)&s=1 \\
\\ \left( 1-\frac{\cos\left[(s-1)\arctan \tau\right]}{\left(1+\tau^2\right)^{\frac{s-1}{2}}}\right)\Gamma[s-1]&s\neq1
\end{array}
\right.
\label{gamtau}
\end{align}
where  $\tau=\omega_c t$ and $\Gamma[x]=\int_0^\infty t^{x-1}e^{-t } dt$ 
is the Euler Gamma function. 
\begin{color}{black}
For short time we have $\gamma_s (\tau)\simeq \frac12 \tau^2 \Gamma[1+s]$ 
$\forall s$, whereas for large value of $\tau$ the dephasing rate 
diverges for $s\leq1$, and shows a finite asymptotic value 
$\gamma_s (\infty) = \Gamma[s-1]$ 
for $s>1$.
In Fig. \ref{f:gamma}, we show the behaviour of $\gamma_s (\tau)$ 
as a function of $\tau$ for three different values of $s$. 
\begin{figure}[h!]
\includegraphics[width=0.9\columnwidth]{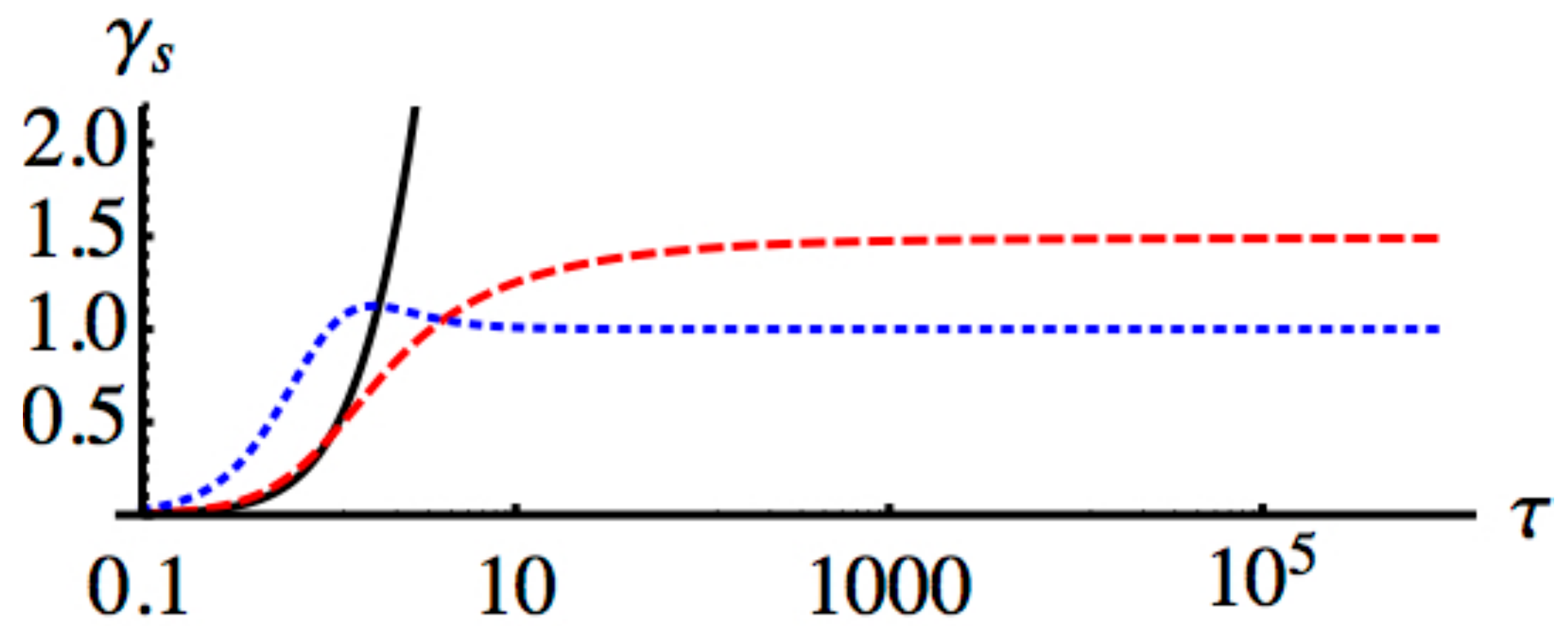}
\includegraphics[width=0.9\columnwidth]{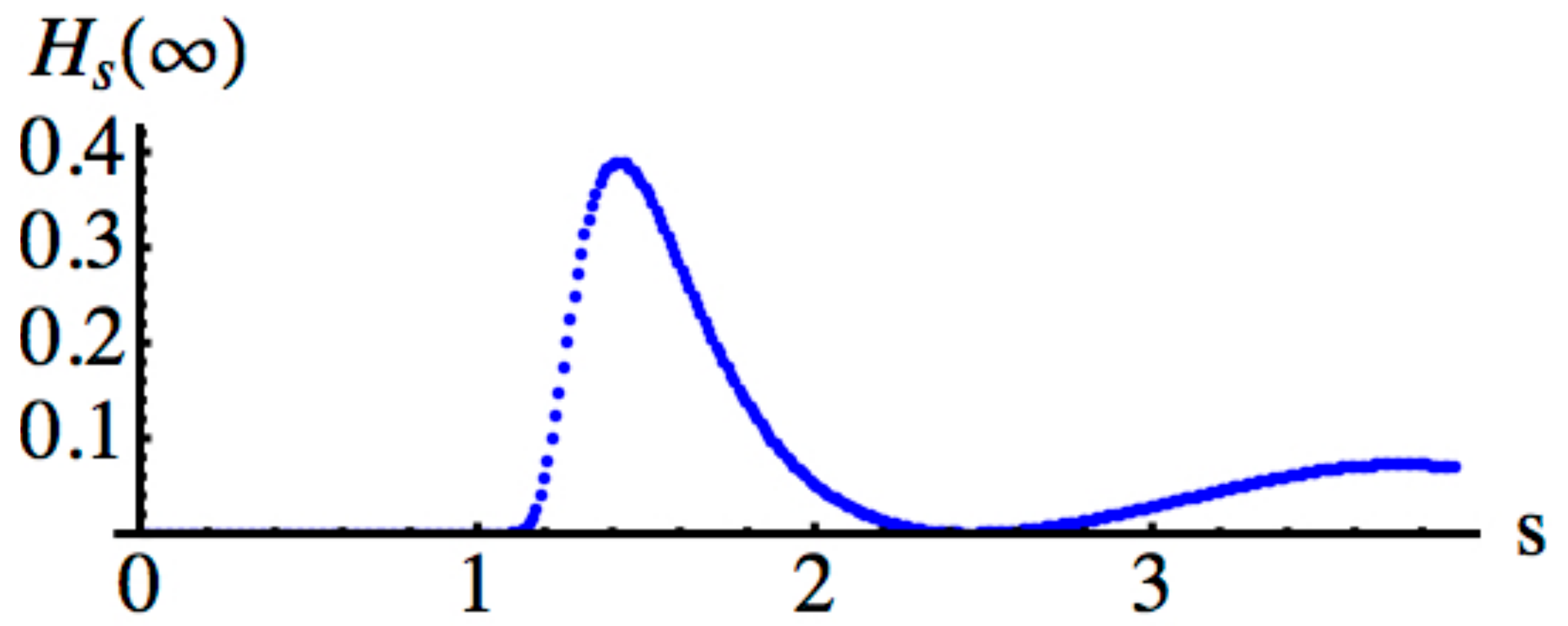}
\begin{color}{black}
\caption{(Top): the dephasing rate $\gamma_s (\tau)$ as a function 
of $\tau$ for three values of $s=0.1$ (solid black line), 
$1.6$ (dashed red line), and $3.0$ (dotted blue line).
(Bottom): the asymptotic value $H_s(\infty)$ of the QFI as a function 
of the ohmicity parameter $s$. We have $H_s(\infty)=0$
for $s\leq 1$, and $H_s(\infty)\neq0$ for $s>1$.
\label{f:gamma}}\end{color}
\end{figure}
\end{color}
\par
Upon preparing the qubit in the initial state $|+\rangle = (|0\rangle + |1\rangle)/\sqrt{2}$, the evolved state $\varrho_{s}^+ (\tau)$ is obtained
from Eqs. (\ref{rhf}) and (\ref{gamtau}). Then,
using Eq. (\ref{cqfi}) the QFI for the estimation of $s$ is given by
\begin{align}
H_s (\tau) 
= \frac{\big[\partial_s \gamma_s (\tau)\big]^2}{e^{2 \gamma_s (\tau)}-1}\,,
\end{align}
where the interaction time $\tau$ is a free parameter that can be used to 
further optimise precision. 
\begin{color}{black}
Notice that the short-time behaviour of 
$H_s (\tau)$, as well as the asymptotic one, may be extracted from the 
corresponding behaviour of the dephasing rate $\gamma_s (\tau)$. 
For short time we have $H_s (\tau)\simeq  \tau^2 g_s$, where
$$g_s = \frac{\Gamma[s-1]}{4s(s-1)} \left(2s-1 + s (s-1) \psi[s-1]\right)^2
\,,$$ $\psi[z]=\Gamma^\prime[z]/\Gamma[z]$ being the logarithmic 
derivative of the gamma function. For large value of $\tau$ we have $H_s(\infty)=0$
for $s\leq1$, and $H_s(\infty)\neq0$ for $s>1$. The behaviour of the
asymptotic value $H_s(\infty)$ as a function of the ohmicity parameter 
is illustrated and summarised in the lower panel of Fig. \ref{f:gamma}.
\end{color}
\par
\begin{figure}[h!]
\begin{center}
\includegraphics[width=0.9\columnwidth]{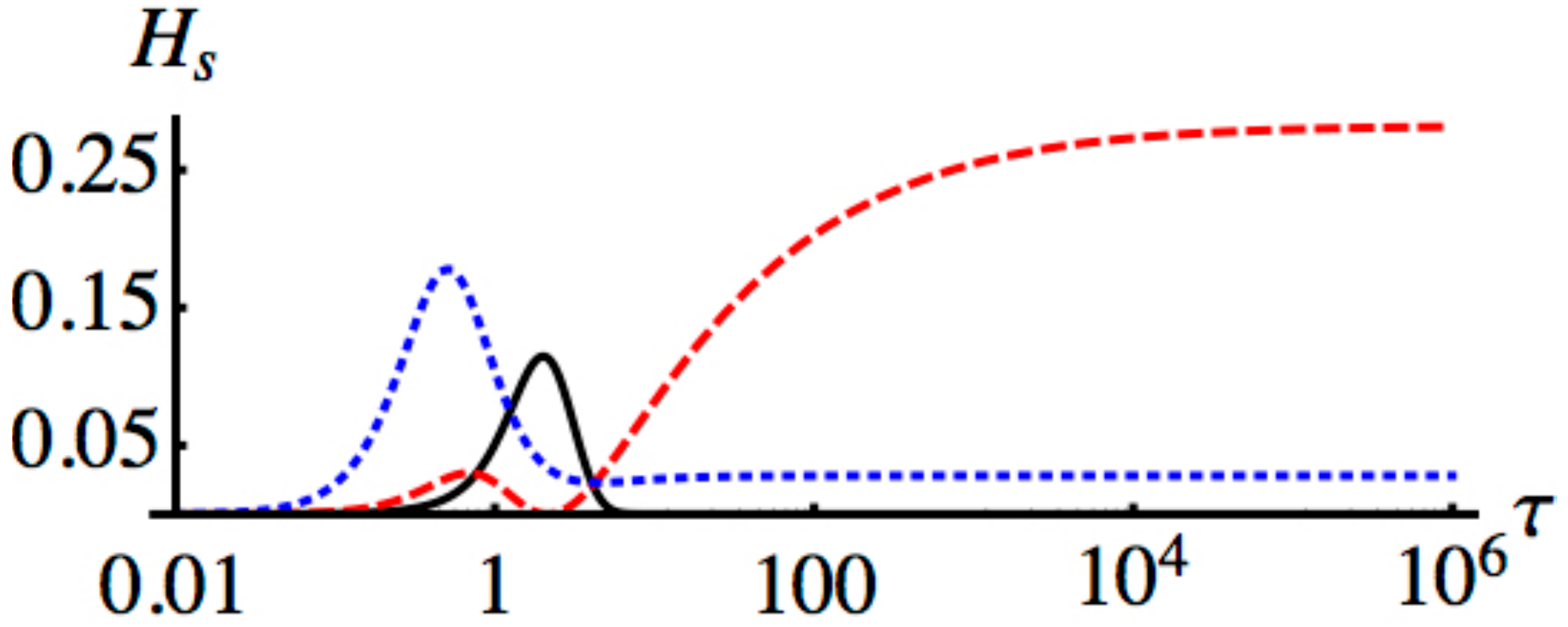}
\includegraphics[width=0.54\columnwidth]{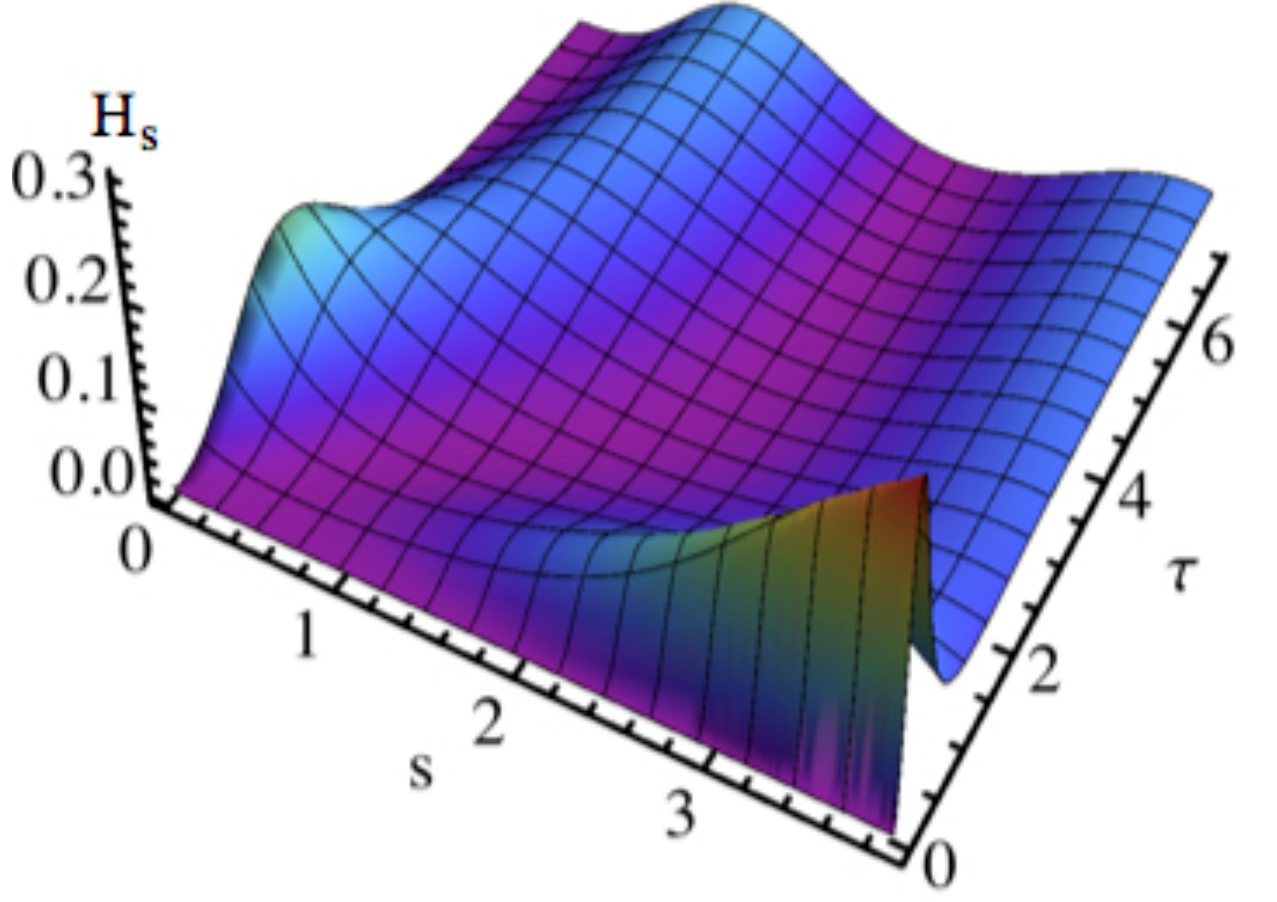}
\includegraphics[width=0.4\columnwidth]{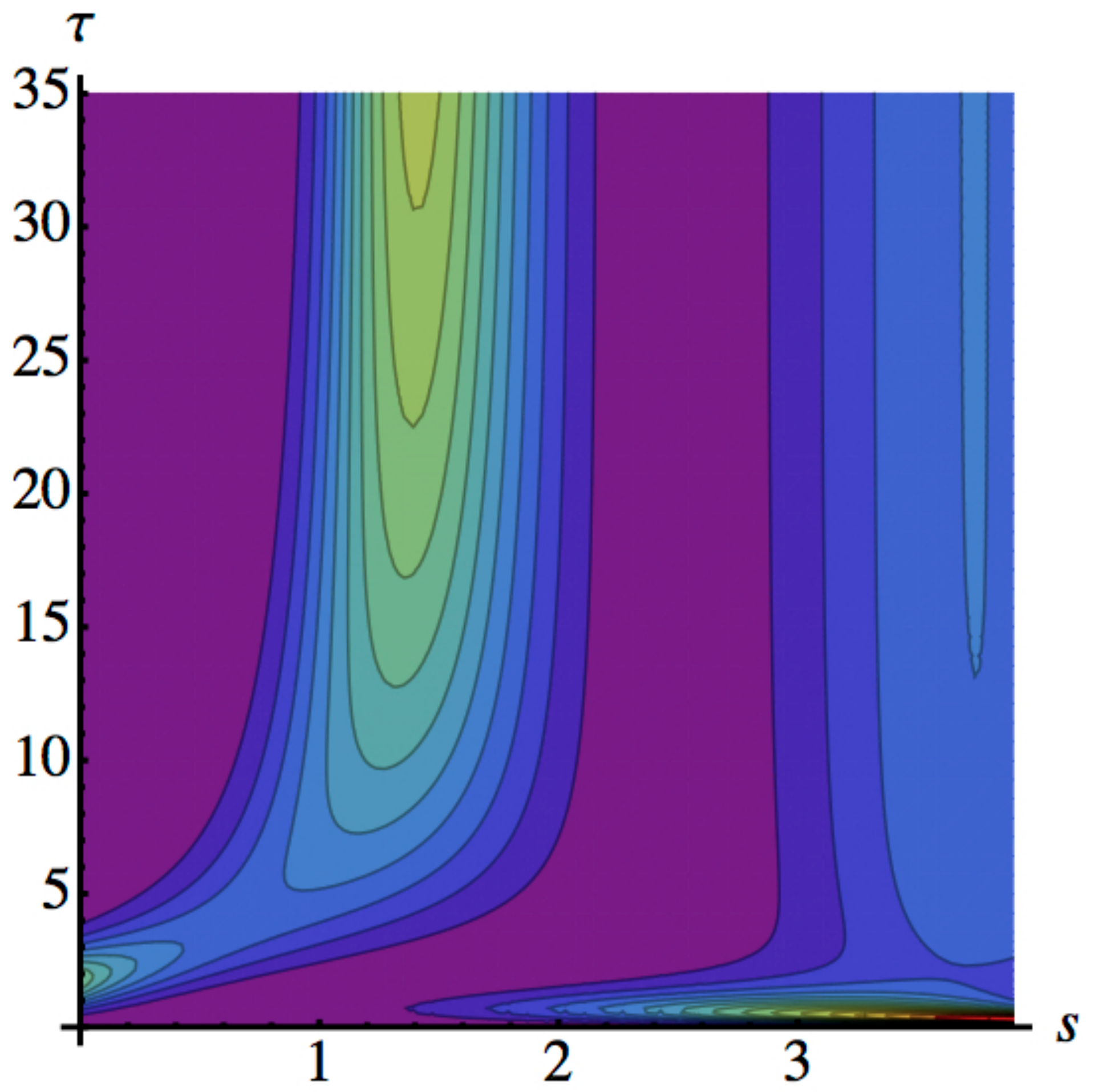}
\includegraphics[width=0.9\columnwidth]{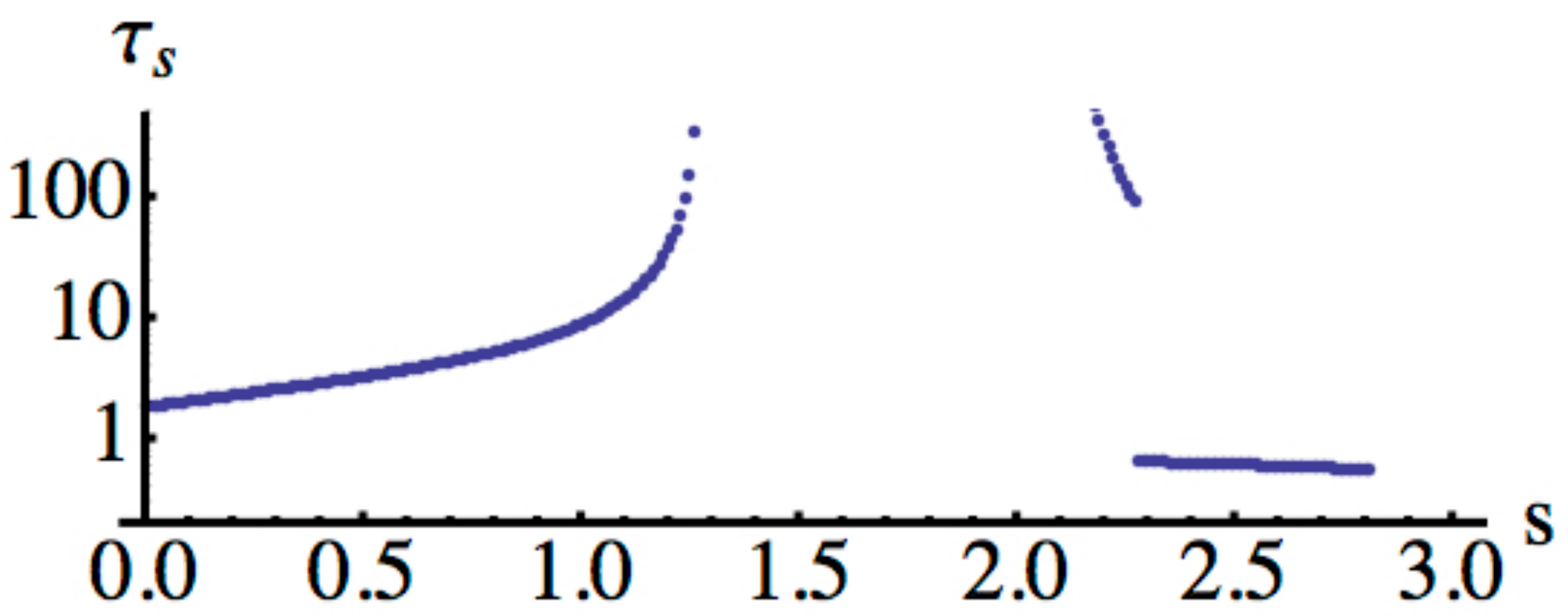}
\end{center}
\caption{(Top): The QFI $H_s$ as a function of $\tau$ 
for three values of $s=0.1$ (solid black line), $1.6$ (dashed red 
line), and $3.0$ (dotted blue line). (Middle): The QFI $H_s$ 
as a function of $s$ and $\tau$. On the left a 
3D plot for interaction times $\tau \leq 7$, and on the right 
a contour plot illustrating the behaviour for a longer time range 
($\tau \leq 35$). (Bottom): The time value $\tau_s$, maximising 
$H_s$ at fixed $s$. The value $\tau_s$ increases with $s$ when $s$ 
is small and then jump to a smaller value for larger values of $s$.
In the intermediate region, $H_s$ is an increasing function of $\tau$.
}\label{f:hst}
\end{figure}

In Fig. \ref{f:hst} we show the behaviour of 
$H_s$ as a function of $s$ and $\tau$. In order to emphasise the non
trivial features of this function, we first show a 2D plot as a function of
time for three values of $s$ (upper panel), a 3D plot illustrating 
the behaviour for limited interaction time ($\tau \leq 7$, middle left 
plot) and a contour plot with a longer time range ($\tau \leq 35$, middle right
plot). In order to optimise the estimation of $s$ we should chose the 
interaction time that maximises the QFI. As it may be guessed from the 
plots, this optimal time $\tau_s$ increases with $s$ when $s$ is small,
and then jump to a smaller value for larger values of $s$. We obtain 
numerically the following estimate $\tau_s \simeq \pi e^s/2 $ for
$s \ll 1$ and $\tau_s=\pi/(2s)$ for $2.2\lesssim s\lesssim 3$. 
\par
In the 
intermediate region, $H_s$ is an increasing (though saturating) function
of $\tau$, and the optimal strategy would be to leave the qubit interacting 
with the environment as much as it can. Of course, this is not possible, 
due to the finite size of any environment. Thus the prescription is that of 
choosing a generically {\em large} interaction time. Overall, we have that
depending on the values of $s$, one may achieve optimal estimation at
finite interaction time (i.e. out-of-equilibrium) or for large  time (when,
presumably, the qubit has reached its stationary state \cite{uqms}).
The behaviour of $\tau_s$ as a function of $s$ is shown in the lower
panel of Fig. \ref{f:hst}.
Our results confirm that pure dephasing is an effective mechanism to 
gain information about the system under investigation without exchanging 
energy.
\par
\begin{figure}[h!]
\begin{tabular}{l}
$\quad$ \includegraphics[width=0.92\columnwidth]{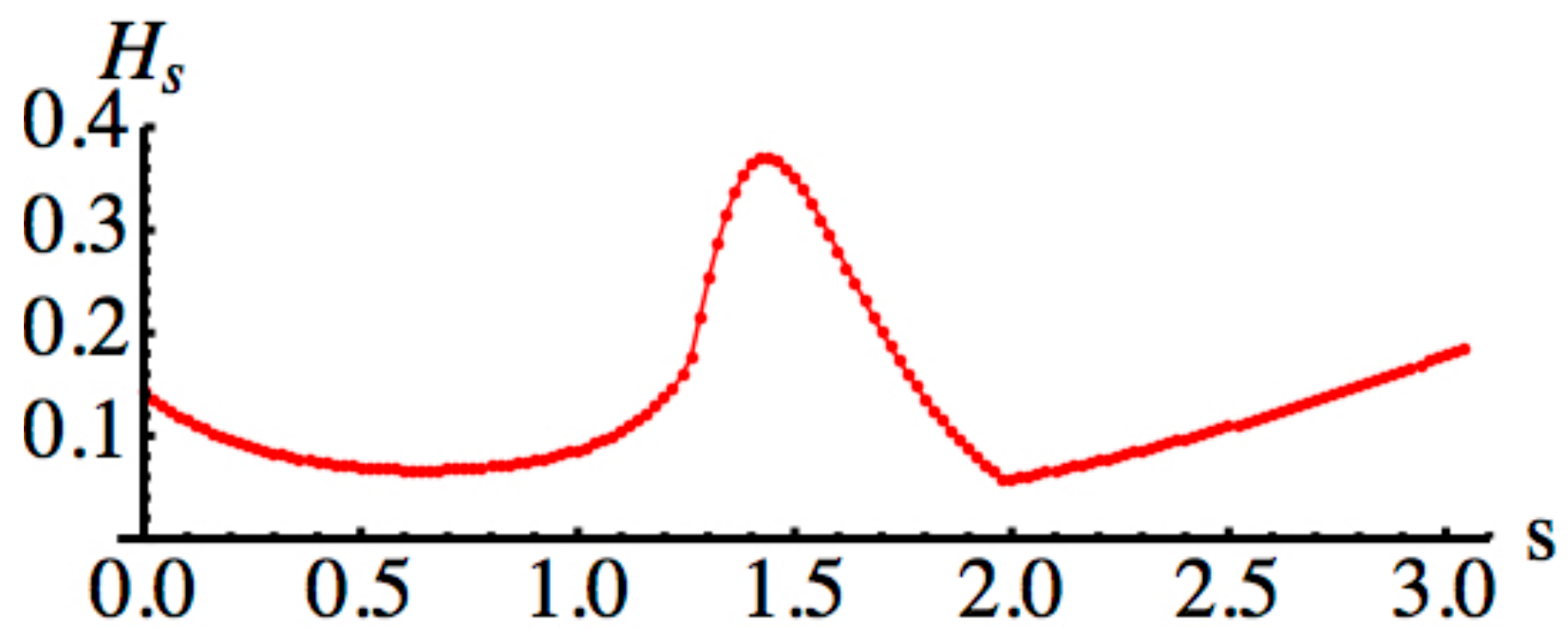} \\
\includegraphics[width=0.95\columnwidth]{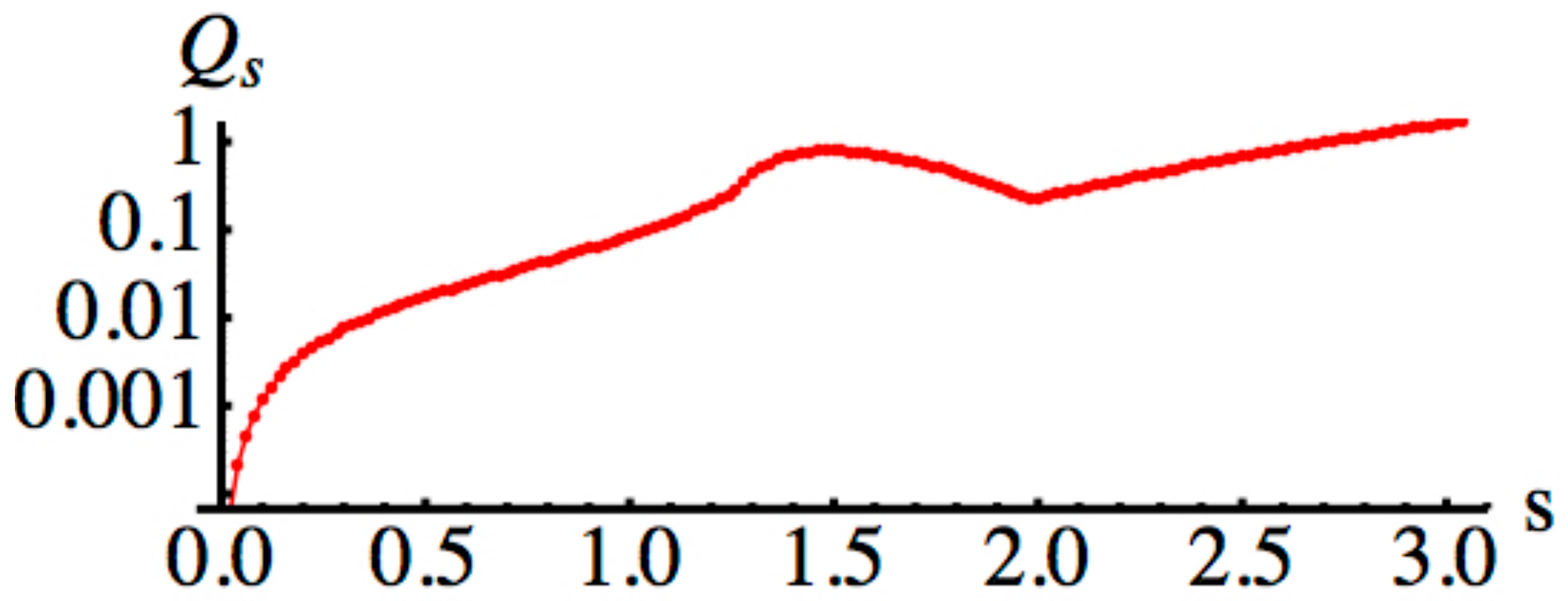}
\end{tabular}
\caption{(Top): The optimal value QFI $H_s$, maximised over $\tau$, as a 
function of $s$. (Bottom): the corresponding quantum 
signal-to-noise ratio $Q_s$. As it is apparent from the plots, 
estimation of intermediate values of $s$, corresponding to 
slightly super-Ohmic environments ($s\simeq 1.5$), is inherently 
more precise than the estimation of smaller values, and also of 
slightly larger values.}\label{f:hopt}
\end{figure}

In the upper panel of Fig. \ref{f:hopt} we show the optimised value of 
$H_s$ as a function of $s$, whereas in the lower panel we show the 
corresponding quantum signal-to-noise ratio $Q_s=s^2 H_s$.
The physical meaning of these plots is that estimation of intermediate 
values of $s$, corresponding to slightly super-Ohmic environments 
($s\simeq 1.5$), is inherently more precise than the estimation 
of smaller or slightly larger values ($1.5\lesssim s \lesssim 2.5$).
\begin{color}{black}
\subsection{Feasible measurement achieving optimal precision}
In this Section we discuss the feasibility of the optimal 
measurement, i.e. whether it exists a measurement for which 
the associated Fisher information is equal to the QFI.
To this aim, let us consider the most general projective
measurement $\{P_\pm\}$, $P_+ +P_-= {\mathbb I}$ on the
qubit probe, i.e. 
\begin{equation}
P_{\pm}=\frac{{\mathbb I}\pm {\boldsymbol b}\cdot{\boldsymbol \sigma}}{2}\,,
\end{equation}
where  ${\boldsymbol b}=(b_{1},b_{2},b_{3}) $, $|{\boldsymbol b}|= 1$ 
and  ${\boldsymbol \sigma} $ is the vector of the Pauli matrices, 
$\sigma_{x},\sigma_{y},\sigma_{z}$. The probability 
distribution of the two outcomes for the 
qubit probe initially prepared in the state 
$|+\rangle$ is given by 
\begin{equation}
p_{\pm}(\tau) = \hbox{Tr} \left[\varrho_{s}^+(\tau)\, P_{\pm} \right] =
\frac{1}{2} \left[1 \pm b_{1} e^{-\gamma_s(\tau)}\right]\,,
\end{equation}
corresponding to a Fisher information
\begin{equation}
F_s=\sum_{k=\pm}\dfrac{\left[\partial_s
p_{k}(\tau)\right]^{2}}{p_{k}(\tau)}\,.
\end{equation}
Starting from the above equation, it is easy to see 
that we have $ F_s = H_s $ if $ b_{1}=1 $, i.e. for 
the measurement of $\sigma_x$ on the qubit. 
This means that measuring $\sigma_x$ provides optimal 
estimation of temperature, provided that an efficient estimator
is employed to process the data. The overall optimal strategy
thus consists in the preparation of the qubit in an eigenstate of 
$\sigma_x$ and the measurement of the same observable 
after the interaction with the environment.
\subsection{Quantum probes at nonzero temperature}
If temperature $T$ of the environment is not strictly zero, the dephasing
rate is given by
\begin{align}
\gamma_s (\tau,T) & =\int_0^{\infty}\!\!\!\!\!d\omega\,  
\frac{1-\cos(\omega \tau/\omega_c)}
{\omega^2}\,J_s(\omega)\, \coth \frac{\omega}{2 T}\,\,, 
\end{align}
which is usually hard to evaluate analytically for a generic value of
$s$. In many situation of interest, however, the temperature is not 
too high and we may use the approximate expression 
$$\coth \frac{\omega}{2 T} \stackrel{T\ll1}{\simeq} 1+2 e^{-\omega/T}\,.$$
In those situations, the dephasing rate may be written as 
\begin{align}\label{tgm1}
\gamma_s (\tau,T) & \simeq \gamma_s (\tau,0) + 
2 \left(\frac{1+T}{T}\right)^{1-s}\!\!\!
\gamma_s \left(\frac{T \tau}{1+T},0\right) \\ \label{tgm2}
& \simeq \gamma_s (\tau,0) + \frac{T^{1+s} (1-T)}{(1+T)^s}\,\tau^2\, 
\Gamma[1+s]\,,
\end{align}
where $\gamma_s (\tau,0)$
is given in Eq. (\ref{gamtau}), and 
temperature is an adimensional quantity, expressed in unit 
of $\omega_c$ \cite{asym}. Using Eq. (\ref{tgm2}), an analytic 
expression for the QFI $H_s$ at low temperature may be 
obtained and compared 
with the zero temperature case. In order to quantify the 
effects of temperature we introduce the excess QFI 
\begin{align}
\Delta H_s (\tau,T) = H_s(\tau,T) - H_s(\tau, T\equiv 0)\,,
\end{align}
which is positive when a nonzero temperature leads to an
improvement in precision, and negative otherwise. In Fig. 
\ref{f:exc} we show the absolute value $|\Delta H_s (\tau,T)|$ 
of the excess QFI as a function of $s$ and $\tau$ for 
two different values of the temperature. On the left we show 
the function for $T=\omega_c/100$ and on the right for 
$T=\omega_c/10$. The blue region corresponds to 
$\Delta H_s (\tau,T)>0$, i.e. working at nonzero temperature is 
convenient in terms of the achievable precision, and the 
green one to $\Delta H_s (\tau,T)<0$, i.e. regions where temperature
is degrading performances. Upon looking at Fig. 
\ref{f:exc} and noticing the different scale with 
respect to  Fig. \ref{f:hst}b, one concludes that (low)
temperature has only a minor effect on the achievable 
precision. Accordingly, the optimal interaction time 
for the probe, and the resulting maximum value of the 
QFI are only slightly changed.
\begin{figure}[h!]
\includegraphics[width=0.45\columnwidth]{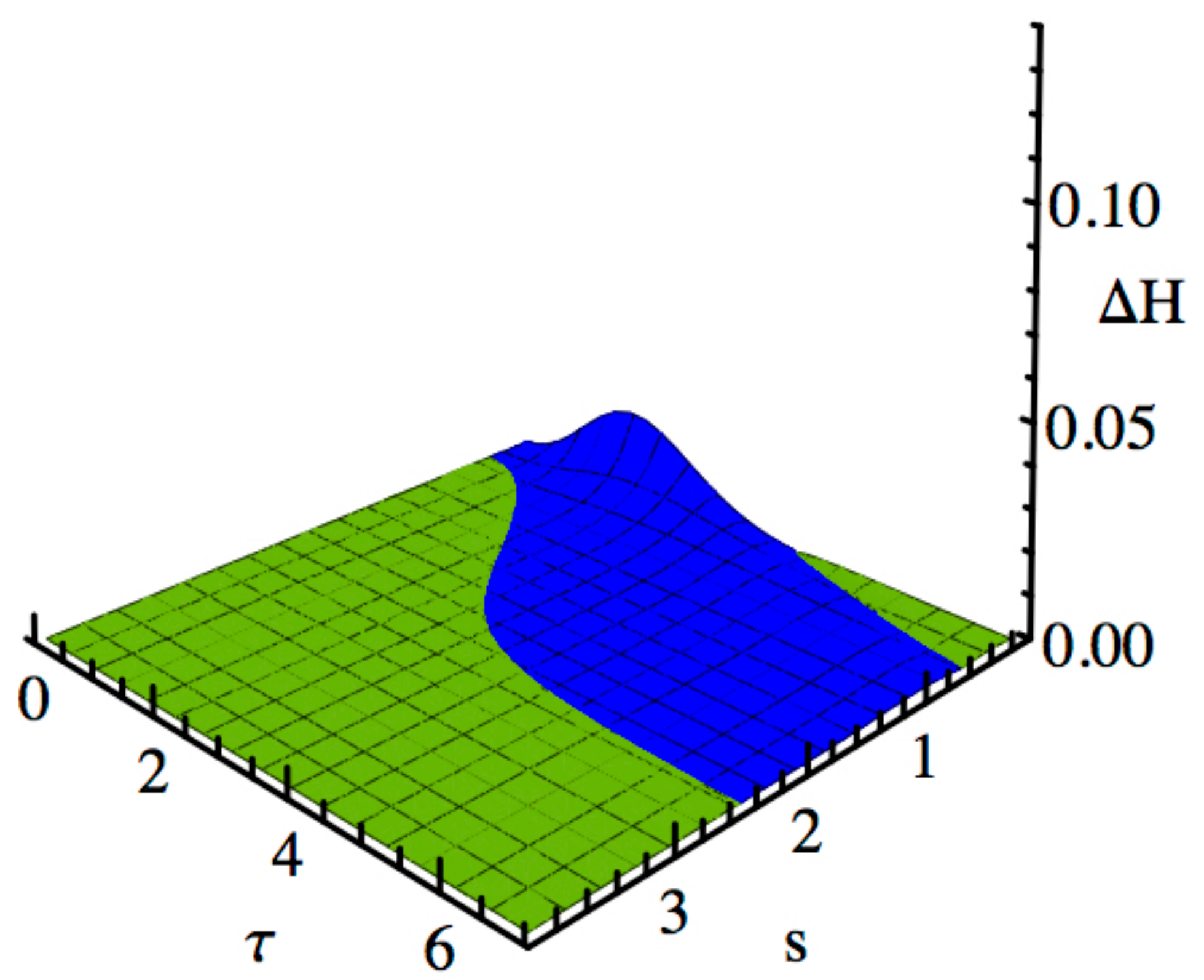}
\includegraphics[width=0.44\columnwidth]{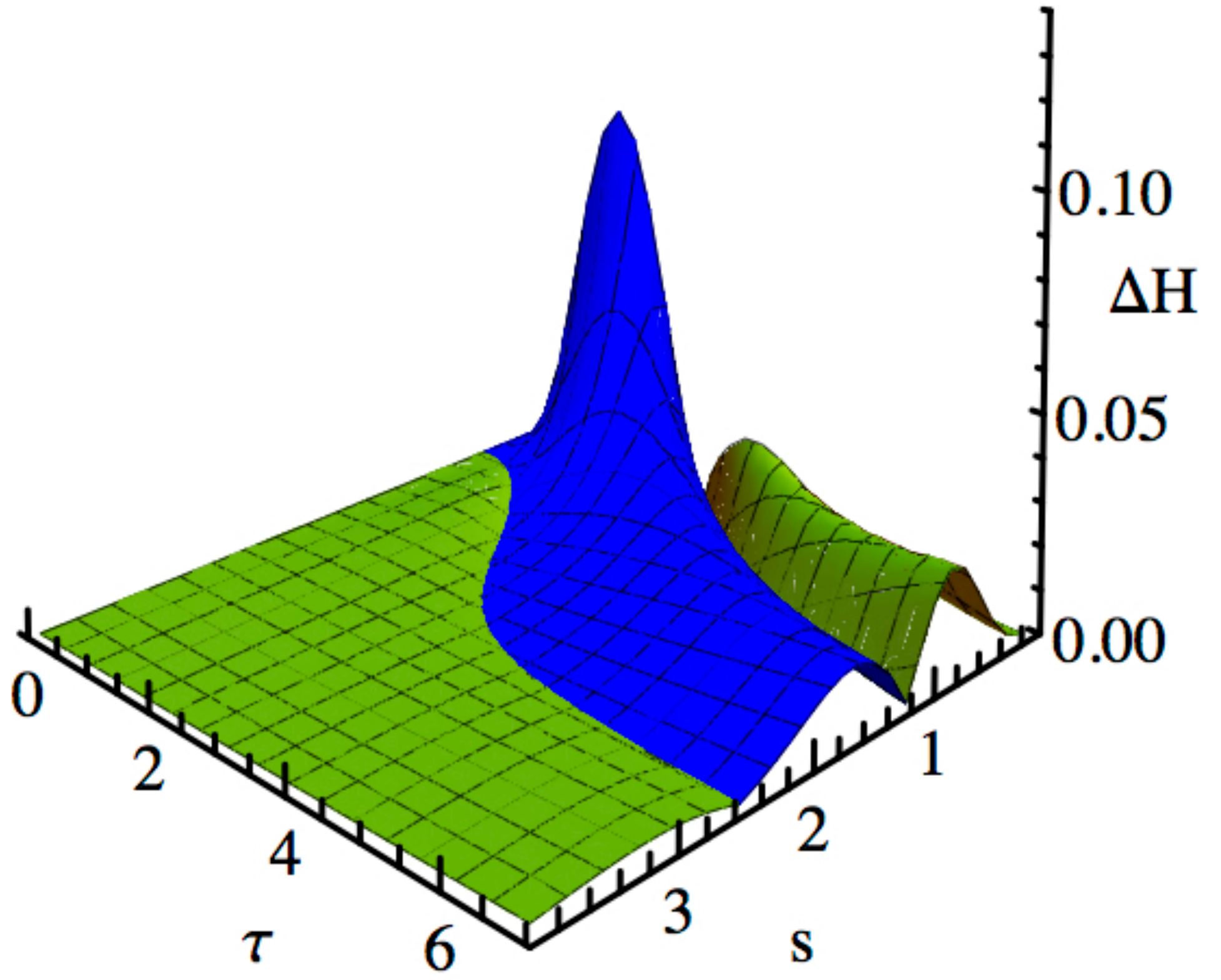}
\begin{color}{black}
\caption{The absolute value $|\Delta H_s(\tau,T)|$ of the excess QFI 
as a function of $s$ and $\tau$ for two different values
of the temperature. On the left the function
for $T=\omega_c/100$ and on the right for $T=\omega_c/10$.
The blue region corresponds to $D_s (\tau,T)>0$ and the green one
to $D_s (\tau,T)<0$. Notice the different scale with respect to 
Fig. \ref{f:hst}b.}\label{f:exc}
\end{color}
\end{figure}
In the opposite limit, i.e. when the temperature is high, 
the situation dramatically changes. This may be easily seen
by expanding the hyperbolic cotangent as $\coth x \simeq x^{-1}$, 
thus arriving at  $$\gamma_s(\tau,T) \simeq 
\frac{1}{2T}\,\gamma_{1+s}(\tau,0)\,.$$
Moreover, since $\gamma_{1+s}(\tau,0)$ is a bounded function 
of $\tau$, we have
\begin{align}
H_s(\tau,T) & = 
\frac{1}{4T^2} \frac{\Big[\partial_s \gamma_{1+s}(\tau,0)\Big ]^2}
{e^{\gamma_{1+s}(\tau,0)/T}-1} \notag \\ 
& \approx \frac1T H_{1+s}(\tau,0)\,. \label{ssp}
\end{align}
Eq. (\ref{ssp}) says that the QFI for the ohmicity parameter
at high temperature is largely reduced in comparison to 
the low temperature case, i.e. almost no information may 
be extracted by quantum probes. Indeed, this is matching 
physical intuition, since for large temperature decoherence 
is mostly due to thermal fluctuations rather than the specific 
features of the interaction, and thus the probes is unable 
to extract information about the structure of the environment.
\end{color}
\section{Conclusions}
In this paper we have addressed the use of open quantum 
systems out-of-equilibrium as possible quantum probes for the characterisation 
of their environment. In particular, we have discussed  
estimation schemes involving parameters governing a de-phasing evolution 
of the probe. For qubit probe we have found a simple relation linking 
the quantum Fisher information to the residual coherence of the probe. 
Finally, we have addressed in some details the estimation of the 
ohmicity parameter of a bosonic environment, finding that depending 
on the values of $s$, one may achieve optimal estimation at
finite interaction time, i.e. when the probe is in an out-of-equilibrium
state, or for large  time, when, presumably, the qubit has reached its 
stationary state. Overall, our results pave the way for 
further investigation in out-of-equlibrium quantum metrology, 
perhaps exploiting memory effects \cite{nm0}, and confirm that 
pure dephasing at low temperature represents an effective mechanism 
to imprint information on quantum probes without 
exchanging energy with the system under investigation.
\section*{Acknowledgements}
This work has been supported by CARIPLO foundation through the 
Lake-of-Como School program, and by SERB through the VAJRA scheme 
(grant VJR/2017/000011). MGAP is member of GNFM-INdAM.
The authors are grateful to Matteo Bina, Claudia Benedetti, 
and Luigi Seveso, for useful discussions.


\begin{thebibliography}{99}
\bibitem{asym} More generally, upon expanding the hyperbolic cotangent as $\coth x = (1+ e^{-2x}) \sum_{n=0}^\infty e^{-2nx}$, we obtain 
$$
\gamma_s (\tau, T) = \gamma_s (\tau,0) + 2 \sum_{n=1}^\infty a_n^{1-s}\, \gamma_s \left(\frac{\tau}{a_n},0\right)\,,
$$
where$$ a_n \equiv a_n (T) = 1+ \frac{n}{T}\,,$$
with $\tau$ and $T$ expressed in unit of $\omega_c$.
\end{thebibliography}
\end{document}